\documentclass[10pt,conference]{IEEEtran}
\usepackage{amsmath,subfigure}
\usepackage{amssymb}
\usepackage{amsthm}
\usepackage{verbatim}
\usepackage{algorithm,algorithmic}
\usepackage{graphicx}
\usepackage{cite}
\usepackage[usenames]{color}

\def\inDim{N} 
\def\outDim{M} 
\def\Rate{{\bd{\lambda}}} 
\def\rate{\lambda} 
\def\Flow{\bd{X}} 
\def\flow{X} 
\def\Cts{\bd{Y}} 
\def\SenMat{\bd{A}} 
\def\res{\tau} 

\def\deq{\triangleq}

\def\R{\mathbb{R}} 
\def\Rplus{\R_+} 
\def\Z{\mathbb{Z}} 
\def\Zplus{\Z_+} 

\def\argmin{\operatornamewithlimits{argmin}}
\def\Prob{{\mathbb P}}
\def\Exp{{\mathbb E}}

\def\cN{{\mathcal N}}

\def\bd#1{\boldsymbol{#1}}

\def\wh#1{\widehat{#1}}

\def\Risk#1#2{R\left( #1, #2\right)} 
\def\kerr#1{\sigma_k(#1)} 
\def\hRate{\wh{\Rate}} 
\def\hrate{\wh{\rate}} 
\def\hFlow{\wh{\Flow}}

\def\Estimators{{\mathcal F}}
\def\dir{{\rm dir}}
\def\pen{{\rm pen}}
\def\pMLE{{\rm pMLE}}
\def\Poi{{\rm Poisson}}

\def\Graph{G}

\def\Edges{E}
\def\Vin{A} 
\def\Vout{B} 
\def\AdjMat{\bd{A}} 
\def\nAdjMat{\bd{\Phi}} 
\def\MinSet{\Omega} 

\def\eps{\varepsilon}
\def\tO{\widetilde{O}}

\newtheorem{theorem}{Theorem}
\newtheorem{definition}{Definition}
\newtheorem{proposition}[theorem]{Proposition}

\renewcommand{\k}{k}
\newcommand{\n}{\inDim}
\newcommand{\m}{\outDim}
\newcommand{\T}{n}

\renewcommand{\a}{\Rate}

\newcommand{\has}{\hRate}

\begin{document}

\title{Fishing in Poisson Streams: Focusing on the Whales, Ignoring the Minnows}

\author{
\IEEEauthorblockN{Maxim Raginsky}
\IEEEauthorblockA{ECE, Duke University \\
Durham, NC 27708, USA}
\and
\IEEEauthorblockN{Sina Jafarpour}
\IEEEauthorblockA{CS, Princeton University
   \\ Princeton, NJ 08540, USA}
\and
\IEEEauthorblockN{Rebecca Willett}
\IEEEauthorblockA{ECE, Duke University\\
Durham, NC 27708, USA}
 \and
 \IEEEauthorblockN{Robert Calderbank}
 \IEEEauthorblockA{EE, Princeton University
    \\ Princeton, NJ 08540, USA}
}

\maketitle
\thispagestyle{empty}


\begin{abstract} This paper describes a low-complexity approach for reconstructing average packet arrival rates and instantaneous packet counts at a router in a communication network, where the arrivals of packets in each flow follow a Poisson process.  Assuming that the rate vector of this Poisson process is sparse or approximately sparse, the goal is to maintain a compressed summary of the process sample paths using a small number of counters, such that at any time it is possible to reconstruct both the total number of packets in each flow and the underlying rate vector. We show that these tasks can be accomplished efficiently and accurately using compressed sensing with expander graphs. In particular, the compressive counts are a linear transformation of the underlying counting process by the adjacency matrix of an unbalanced expander. Such a matrix is binary and sparse, which allows for efficient incrementing when new packets arrive. We describe, analyze, and compare two methods that can be used to estimate both the current vector of total packet counts and the underlying vector of arrival rates.
\end{abstract}

\section{Introduction}

Successful management of large-scale communication networks rests crucially on the availability of accurate traffic measurements. From the viewpoint of such  tasks as billing/accounting or intrusion detection, network traffic is composed of packet flows (or streams) arriving at or departing from routers in the network. As both the number of users and the data rates continue growing, there is increasing emphasis on traffic measurement architectures that are accurate, fast, and cheap. Naturally, some trade-offs between these three desiderata are inevitable. For instance, one could keep a dedicated counter for each flow, but, depending on the type of memory used, one could end up with an implementation that is either fast but expensive and unable to keep track of a large number of flows (e.g.,~using SRAMs, which have low access times, but are expensive and physically large) or cheap and high-density but slow (e.g.,~using DRAMs, which are cheap and small, but have longer access times).

Recent work has shown that a reasonable compromise between accuracy, speed and cost can be found if one takes into account certain prior knowledge about the relative flow sizes in a typical network. In particular, there is empirical evidence \cite{EmpiricalFlow1,EmpiricalFlow2} that flow sizes in IP networks follow a {\em heavy-tail} pattern: just a few flows (say, $10\%$) carry most of the traffic (say, $90\%$). Based on this observation, Estan and Varghese \cite{ElephantsMice} proposed two methodologies (``sample-and-hold" and ``multistage filters") that use a small number of counters to keep track only of the flows whose sizes exceed a given fraction of the total bandwidth. More recently, Lu et al.~\cite{CounterBraids} developed a new technique, termed ``Counter Braids," which uses sparse random graphs to aggregate (or ``braid") the raw packet counts into a small number of counters. The total size of {\em each} flow can then be recovered at the end of a measurement epoch using a message passing decoder.

The approach of Estan and Varghese \cite{ElephantsMice} allows one to keep track only of the few heavy flows while ignoring the rest (``focusing on the elephants, ignoring the mice," as they put it), while Lu et al.~\cite{CounterBraids} can recover the {\em entire} vector of flow sizes. Moreover, these two approaches rely on different modeling assumptions. Specifically, in \cite{ElephantsMice} the flow sizes are assumed to be deterministic and subject to the heavy-tail behavior, while in \cite{CounterBraids} the flow sizes are i.i.d.\ realizations of a random variable with a heavy-tail distribution.

\subsection{Our contribution}

The present paper considers a more realistic setting where each flow (or stream) is modeled as a Poisson process with an unknown rate (measured in packets per unit time), and it is the {\em rates} corresponding to the streams at a given router that possess the heavy-tail property. This modeling assumption combines certain aspects of \cite{ElephantsMice} and \cite{CounterBraids}: the heavy-tail property is present both on the level of coarse-grained, time-averaged behavior of the flows and on the level of actual traffic patterns, which are stochastic. Moreover, our model goes beyond the i.i.d.\ assumption of \cite{CounterBraids} and can account for the heterogeneous nature of the different flows entering a particular router.

The main goal is to reconstruct the underlying vector of rates while maintaining a small number of counters with low access times. To accomplish this goal, we exploit our recent work \cite{Asilomar} on compressed sensing (CS) with Poisson-distributed observations. Mathematically, the heavy-tail property can be restated as follows: the vector of rates is, to a good approximation, {\em sparse}. This sparsity interpretation strongly suggests that CS can be used to accurately recover the underlying vector of rates from a small number of judiciously designed linear transformations of the observed flows. Building on the results from \cite{Asilomar}, we show that the raw packet counts can be mapped into a small number of ``compressed" counts using the adjacency matrix of a properly constructed unbalanced expander. Such an adjacency matrix has binary entries and is sparse (i.e.,~each column has a small constant number of ones), which ensures that the counts can be updated using a small number of operations as new packets arrive. The resulting architecture can be used to recover the raw packet counts as well. Since we are dealing here with Poisson streams, we would like to push the metaphor further and say that we are ``focusing on the whales, ignoring the minnows."

We analyze the performance of our scheme theoretically, describe an efficient implementation, and present preliminary experimental results.

 \subsection{Notation}
 
 Given a vector $\bd{u} \in \R^m$ and a set $S \subseteq \{1,\ldots,m\}$, we will denote by $\bd{u}^S$ the vector obtained by setting to zero all coordinates of $\bd{u}$ that are in $S^c$, the complement of $S$: $\forall 1 \le i \le m, u^S_i = u_i 1_{\{ i \in S \}}$. Given some $1 \le k \le m$, let $S$ be the set of positions of the $k$ largest (in magnitude) coordinates of $\bd{u}$. Then $\bd{u}^{(k)} \deq \bd{u}^S$ will denote the {\em best $k$-term approximation} of $\bd{u}$ (in any norm on $\R^m$), and 
$$
\kerr{\bd{u}} \deq \| \bd{u} - \bd{u}^{(k)} \|_1 = \sum_{i \in S^c} |u_i|
$$
will denote the resulting $\ell_1$ approximation error. The $\ell_0$ quasinorm measures the number of nonzero coordinates of $\bd{u}$: $\| \bd{u} \|_0 \deq \sum^m_{i=1} 1_{\{ u_i \neq 0 \}}$. Given a vector $\bd{u}$, we will denote by $\bd{u}^+$ the vector obtained by setting to zero all negative components of $\bd{u}$: for all $1 \le i \le m$, $u^+_i = \max\{0,u_i\}$.
 
 \section{Problem formulation}
 
 We wish to monitor a large number $\inDim$ of packet flows using a much smaller number $\outDim$ of counters. Each flow is a homogeneous Poisson process. Specifically, let $\Rate^\star \in \Rplus^\inDim$ denote the vector of rates, and let $\bd{U}$ denote the random process
 $\bd{U} = \{\bd{U}_t\}_{t \in \Rplus}$ with sample paths in $\Zplus^\inDim$, where for any $t \in \Rplus$ and any $\bd{k} \in \Zplus^\inDim$ we have
 \begin{align*}
 \Prob_{\Rate^\star}(\bd{U}_t = \bd{k}) = \prod^\inDim_{i=1} \frac{(\rate^\star_i t)^{k_i}}{k_i!} e^{-t\rate^\star_i}.
 \end{align*}
 In other words, for each $i \in \{1,\ldots,\inDim\}$, the $i$th component of $\bd{U}$, which we will denote by $U^{(i)}$, is a homogeneous Poisson process with the rate of $\rate_i^\star$ arrivals per unit time, and all the $U^{(i)}$'s are mutually conditionally independent given $\Rate^\star$.
 
The counters are updated in discrete time, every $\res$ time units. Let $\Flow = \{\Flow_n\}_{n \in \Zplus}$ denote the sampled version of $\bd{U}$, where $\Flow_n \deq \bd{U}_{n\res}$. The update takes place as follows. We have a binary matrix $\SenMat \in \{0,1\}^{\outDim \times \inDim}$, and at each time $n$ let $\Cts_n = \SenMat\Flow_n$. The probabilistic law governing the evolution of the counter contents is now
\begin{align*}
\Prob_{\SenMat,\Rate^\star}(\Cts_n = \bd{\ell}) = \prod^\outDim_{j=1} \frac{(n\res \SenMat \Rate^\star)^{\ell_j}_j}{\ell_j !} e^{-(n\res \SenMat\Rate^\star)_j},
\end{align*}
where the case $(\SenMat \Rate^\star)_j = 0$ for any $1 \le j \le \outDim$ is handled using the fact that $\rate^\ell e^{-\rate}/\ell! \to 1_{\{\ell = 0\}}$ as $\rate \to 0$. In other words, $\Cts$ is a sampled version of an $M$-dimensional homogeneous Poisson process with the rate vector $\SenMat \Rate^\star$.

The goal is to estimate the unknown rate vector $\Rate^\star$ after $n$ time steps given $\Cts^n \deq (\Cts_1,\ldots,\Cts_n)$ using an estimator $\hRate_n$ based on $\Cts^n$: $\hRate_n = \hRate_n(\Cts^n)$. We measure the quality of such an estimator by the expected $\ell_1$ risk:
\begin{align*}
\Risk{\hRate_n}{\Rate^\star} \deq \Exp_{\Rate^\star}\left\| \hRate_n - \Rate^\star \right\|_1 = \Exp_{\Rate^\star}\left[\sum^N_{i=1} \left| \hrate_{n,i} - \rate^\star_i \right|\right],
\end{align*}
where the expectation is taken w.r.t.\ the underlying flow process $\Flow \sim \Prob_{\Rate^\star}$. Assuming that the unknown rate vector $\Rate^\star$ is a member of a given class $\Lambda^\star$, we would like to design the counter update matrix $\SenMat$ and an accompanying sequence of estimators $\{ \hRate_n \}$ to attain low risk $\Risk{\hRate_n}{\Rate^\star}$ over $\Lambda^\star$. One particular class of interest, which pertains to the heavy-tail behavior of network traffic, is defined by
\begin{align}\label{eq:heavy_tail}
\Sigma_{\alpha,L_0} \deq \left\{ \Rate \in \Rplus^\inDim : \| \Rate \|_1 \le L_0; \kerr{\Rate} \le L_0 k^{-\alpha}, \forall k  \right\}
\end{align}
for some $L_0 > 0$ and $\alpha \ge 1$. Here, $\alpha$ is the power-law exponent that controls the tail behavior; in particular, the extreme regime $\alpha \to + \infty$ describes the fully sparse setting.

\section{Preliminaries}
\label{sec:prelims}

As we show in the sequel, a good choice for the counter update matrix $\SenMat$ is the adjacency matrix of a suitably constructed expander.  Adjacency matrices of high-quality expanders have been proposed as an alternative to dense, random measurement matrices for sparse signal recovery \cite{BerindeExpanders,BIR,Sina,Khaj,ssmp}. This section summarizes the key results on expanders, as well as the results from our earlier work \cite{Asilomar} on the use of expanders for sparse recovery under the Poisson observation model.

\subsection{Expanders and sparse recovery}

\begin{definition} A {\em $(k,\eps)$-unbalanced expander}, or simply a {\em $(k,\eps)$-expander}, is a bipartite simple graph $\Graph = (\Vin,\Vout,\Edges)$ with left degree $d$, such that for any $S \subset \Vin$ with $|S| \le k$, the set of neighbors $\cN(S)$ of $S$ has size $|\cN(S)| \ge (1-\eps) d |S|$.
\end{definition}

\noindent Here, $\Vin$ (resp.,~$\Vout$) corresponds to the components of the original signal (resp., its compressed representation). Hence, for a given $|\Vin|$, a ``high-quality" expander should have $|\Vout|$, $d$, and $\eps$ as small as possible, while  $k$ should be as close as possible to $|\Vout|$. The following proposition (cf.~\cite{BerindeExpanders,Asilomar}) tells us what we can expect:

\begin{proposition} For any $1 \le k \le N/2$ and any $\eps \in (0,1)$, there exists a $(k,\eps)$-expander with left set size $\inDim$, left degree $d = O\left(\frac{\log(\inDim/k)}{\eps}\right)$ and right set size $\outDim = O\left( \frac{k \log (\inDim/k)}{\eps^2}\right)$.
\end{proposition}

\noindent From this point on, given $\inDim$ and $1 \le k \le \inDim/2$, we will denote by $\Graph_{k,\inDim}$ a fixed expander with $\eps = 1/16$ whose existence is guaranteed by the above proposition (the value of $\eps$ is fixed for convenience). The following proposition is key to the use of expanders for sparse recovery:

\begin{proposition}\label{prop:expander1} Let $\nAdjMat = \AdjMat/d$ be the normalized adjacency matrix of $\Graph_{k,\inDim}$, and let $\bd{u},\bd{v}$ be two vectors in $\R^\inDim$, such that $\| \bd{u} \|_1 \ge \| \bd{v} \|_1 - \Delta$ for some $\Delta > 0$. Then
\begin{align*}
 \| \bd{u} - \bd{v} \|_1 \le 4 \kerr{\bd{u}} + 4 \| \nAdjMat \bd{u} -  \nAdjMat\bd{v} \|_1 + 2 \Delta.
\end{align*}
\end{proposition}

\noindent For future reference, we note that, since our expander is regular, there exists a minimal set $\MinSet \subset \Vin$ of size $\outDim$, such that its neighborhood covers all of $\Vout$, i.e., $\cN(\MinSet) = \Vout$. Let $\bd{I}_\MinSet \in \R^\inDim$ be the vector with components $I_{\MinSet,i} = 1_{\{ i \in \MinSet\}}$. In that case, note that $\nAdjMat \bd{I}_\MinSet \succeq I_{\outDim \times 1}/d$.

\subsection{Expander-based CS under the Poisson model}
\label{ssec:expander_CS_Poisson}

In \cite{Asilomar}, we have considered the following problem: Let $\bd{\theta}^\star \in \Rplus^\inDim$ be an unknown vector of Poisson intensities with known $\ell_1$ norm $\| \bd{\theta}^\star \|_1 = L$ (in general, $L$ may be a known upper bound on $\| \bd{\theta}^\star \|_1$). Given a fixed $1 \le k \le \inDim/2$, let $\nAdjMat$ be the normalized adjacency matrix of $\Graph_{k,\inDim}$. We observe a random vector $\bd{Z} \in \Zplus^\outDim$ distributed according to $\bd{Z} \sim \Poi(\nAdjMat \bd{\theta}^\star)$. 

Let $\Theta \subset \Rplus^N$ be a finite or countable set of candidate estimators of $\bd{\theta}^\star$ such that $\| \bd{\theta} \|_1 \le  L,  \forall \bd{\theta} \in \Theta$, and for a given $c > 0$ define the set
\begin{align*}
\Estimators_c \deq \left\{ \bd{f} \deq \bd{\theta} + c L \bd{I}_\MinSet : \bd{\theta} \in \Theta \right\}.
\end{align*}
Moreover, let $\pen(\cdot) : \Theta \to \Rplus$ be a {\em penalty} (or {\em regularization}) functional satisfying the {\em Kraft inequality},
\begin{align*}
\sum_{\bd{\theta} \in \Theta} e^{-\pen(\bd{\theta})} \le 1.
\end{align*}
Since there is a one-to-one correspondence between $\Theta$ and $\Estimators_c$, we will overload our notation and let $\pen(\bd{f})$ denote $\pen(\bd{\theta})$ whenever $\bd{f} = \bd{\theta} + c L \bd{I}_\MinSet$. In \cite{Asilomar}, we have shown the following:

\begin{proposition}\label{prop:expander2} Consider the {\em penalized maximum likeilhood estimator (pMLE)}
\begin{subequations}\label{eq:PMLE}
\begin{align}
\wh{\bd{f}} &\deq \argmin_{\bd{f} \in \Estimators_c} \left[ - \log \Prob_{\nAdjMat \bd{f}} (\bd{Z}) + 2 \pen(\bd{f}) \right] \label{eq:est_st1} \\
\wh{\bd{\theta}} &\deq \wh{\bd{f}} - c L\bd{I}_\MinSet. \label{eq:est_st2}
\end{align}
\end{subequations}
Then
\begin{align}\label{eq:pMLEbound}
&\Exp \| \bd{\theta}^\star - \wh{\bd{\theta}} \|^2_1 = O \Bigg(\kerr{\bd{\theta}^\star}^2 + c(\outDim L)^2(2d + \outDim c) \Bigg)\nonumber\\
& \,\, + O\Bigg( \left(\frac{d}{c} + M \right) \min_{\bd{\theta} \in \Theta} \left[ \| \bd{\theta}^\star - \bd{\theta} \|^2_1 + \frac{Lc \pen(\bd{\theta})}{d} \right] \Bigg).
\end{align}
\end{proposition}
\noindent Prop.~\ref{prop:expander2} effectively states that the squared $\ell_1$ error of $\wh{\bd{\theta}}$ scales with $M$ times the best penalized $\ell_1$ approximation error plus the $k$-term approximation error of $\bd{\theta}^\star$. The first term in \eqref{eq:pMLEbound} is smaller for sparser $\bd{\theta}^\star$, and the second term is smaller when there is a $\bd{\theta}$ which is simultaneously a good approximation to $\bd{\theta}^\star$ and has a low penalty.

\section{Two estimation strategies}

We consider two estimation strategies. In both cases, we let our measurement matrix $\AdjMat$ be the adjacency matrix of a $\Graph_{k,\inDim}$ for a fixed $k \le \inDim/2$. The first strategy, which we call the {\em direct method}, uses expander-based CS to first recover an estimate of $\Flow_n$ from $\Cts_n$, then constructs an estimate of $\Rate^\star$. The second strategy, which we call the {\em penalized MLE} strategy (or pMLE), relies on the Poisson CS machinery presented in Section~\ref{ssec:expander_CS_Poisson} and can be used when only the rates are of interest. One benefit of pMLE compared to the direct method is its low complexity, which is derived from a preprocessing step based on the structure of the underlying expander $\Graph_{k,\inDim}$.

\subsection{The direct method}

The first approach is to use expander-based CS to obtain an estimate $\hFlow_n$ of $\Flow_n$ from $\Cts_n = \SenMat \Flow_n$, followed by letting
\begin{align}\label{eq:direct}
\hrate_n^\dir = \frac{\hFlow^+_n}{n\res}.
\end{align}
This strategy is based on the observation that $\Flow_n/(n\res)$ is the maximum-likelihood estimator of $\Rate^\star$, and will serve as a ``baseline" against which the penalized MLE will be compared. To obtain $\hFlow_n$, we need to solve the convex program
\begin{align*}
\text{minimize } & \| \bd{u} \|_1 \qquad \text{ subject to }  \AdjMat \bd{u} = \Cts_n
\end{align*}
which can be cast as a linear program \cite{BerindeExpanders}. The resulting solution $\hFlow_n$ may have negative coordinates\footnote{Khajehnejad et al.~\cite{Khaj} have recently proposed the use of perturbed adjacency matrices of expanders to recover nonnegative sparse signals.}, hence the use of the $(\cdot)^+$ operation in \eqref{eq:direct}. We then have the following result:

\begin{theorem}\label{thm:direct}
\begin{align}
\Risk{\hRate_n^\dir}{\Rate^\star} \le 4 \kerr{\Rate^\star} + \frac{ \| (\Rate^\star)^{1/2} \|_1}{\sqrt{n\res}},
\end{align}
where $(\Rate^\star)^{1/2}$ is the vector with components $\sqrt{\rate^\star_i}, \forall i$.
\end{theorem}

\begin{IEEEproof} We first observe that, by construction, $\hFlow_n$ satisfies the relations $\AdjMat \hFlow_n = \AdjMat \Flow_n$ and $\| \hFlow_n \|_1 \le \| \Flow_n \|_1$. Hence,
\begin{align}
\Exp \| \hFlow_n - n\res \Rate^\star \|_1 &\le \Exp \| \hFlow_n - \Flow_n \|_1 + \Exp \| \Flow_n - n \res \Rate^\star \|_1 \nonumber \\
& \le 4 \Exp \kerr{\Flow_n} +  \Exp \| \Flow_n - n \res \Rate^\star \|_1 \label{eq:1s02}
\end{align}
where the first step uses the triangle inequality, while the second step uses Proposition~\ref{prop:expander1} with $\Delta = 0$. To bound the first term in \eqref{eq:1s02}, let $S \subset \{1,\ldots,N\}$ denote the positions of the $k$ largest entries of $\Rate^\star$. Then, by definition of the best $k$-term representation,
\begin{align*}
\kerr{\Flow_n} \le \| \Flow_n - \Flow^S_n \|_1 = \sum_{i \in S^c} |\flow_{n,i}| = \sum_{i \in S^c} \flow_{n,i}.
\end{align*}
Therefore,
\begin{align*}
\Exp \kerr{\Flow_n} \le \Exp \left[ \sum_{i \in S^c} \flow_{n,i} \right] = n\res \sum_{i \in S^c} \rate_i^\star \equiv n \res \kerr{\Rate^\star}.
\end{align*}
To bound the second term, we can use concavity of the square root, as well as the fact that each $\flow_{n,i} \sim \Poi(n\res\rate^\star_i)$, to write
\begin{align*}
 \Exp \| \Flow_n - n \res \Rate^\star \|_1 
  \le \sum^N_{i=1} \sqrt{ \Exp(\flow_{n,i} - n \res \rate^\star_i)^2 } =  \sum^N_{i=1} \sqrt{n\res\rate^\star_i}.
\end{align*}
Now, it is not hard to show that $\| \hFlow^+_n - n\res\rate^\star \|_1 \le \| \hFlow_n - n \res\rate^\star \|_1$. Therefore,
\begin{align*}
\Risk{\hRate_n^\dir}{\Rate^\star} &\le \frac{ \Exp \| \hFlow_n - n\res \Rate^\star \|_1}{n\res} \le 4 \kerr{\Rate^\star} + \frac{ \| (\Rate^\star)^{1/2} \|_1}{\sqrt{n\res}},
\end{align*}
which proves the theorem.
\end{IEEEproof}

\subsection{The penalized MLE approach}

The second approach is based on the penalized MLE framework. Assume that we know a good upper bound $L_0$ on the total average arrival rate $\| \Rate^\star\|_1$. Let $\Lambda$ be a sufficiently large finite set of candidate estimators with $\| \Rate \|_1 \le L_0$ for all $\Rate \in \Lambda$, and let $\pen(\cdot)$ be a penalty functional satisfying the Kraft inequality over $\Lambda$. Given $n$ and $\res$, let $\Lambda_{n,\res} \deq n\res d \Lambda$ with the same penalty function.

We can now apply the results of Section~\ref{ssec:expander_CS_Poisson} with $\bd{Z} = \Cts_n$ and $\bd{\theta}^\star = n\res d \Rate^\star$. With this notation, define
$$
\hRate^\pMLE_n \deq \frac{\wh{\bd{\theta}}}{n\res d},
$$
where $\wh{\bd{\theta}}$ is the corresponding pMLE estimator. Then we have the following risk bound:

\begin{theorem}\label{thm:pMLE} Let $c = \frac{\gamma}{k \log(\inDim/k)}$, where $\gamma >0$ is chosen so that $c \ll 1$. Then
\begin{align}
&\Risk{\hRate^\pMLE_n}{\Rate^\star} = O\left(\kerr{\Rate^\star} +\sqrt{\gamma k} \log (\inDim/k) \right) \nonumber\\
&\qquad + O\left( \log(\inDim/k)  \sqrt{\frac{k}{\gamma}\min_{\Rate \in \Lambda} \left[ \| \Rate^\star - \Rate \|^2_1 + \frac{\pen(\Rate)}{n\res} \right]}\right).
\label{eq:pMLE_bound}
\end{align}
\end{theorem}
We now develop risk bounds under the heavy-tail condition. To this end, let us suppose that $\Rate^\star$ is a member of the heavy-tail class $\Sigma_{L_0,\alpha}$ defined in \eqref{eq:heavy_tail}. Fix a small positive number $\delta$, such that $L_0/\sqrt{\delta}$ is an integer, and define the set
\begin{align*}
 \Lambda \deq \left\{ \Rate \in \Rplus^\inDim: \| \Rate \|_1 \le L_0;  \rate_i \in \{m\sqrt{\delta}\}^{L_0/\sqrt{\delta}}_{m=0}, \forall i \right\}
\end{align*}
These will be our candidate estimators of $\Rate^\star$. We can define the penalty function $\pen(\Rate) \asymp \| \Rate \|_0 \log (\delta^{-1})$ so that it satisfies Kraft's inequality. Moreover, if $\delta$ is small enough, for any $\Rate \in \Sigma_{\alpha,L_0}$ and any $1 \le m \le \inDim$ we will be able to find some $\Rate^{(m)} \in \Lambda$, such that $\| \Rate \|_0 \asymp m$ and
$$
\| \Rate - \Rate^{(m)} \|^2_1 \asymp m^{-2\alpha} + m\delta.
$$
We will also assume that $\delta$ is sufficiently small, so that the penalty term $\frac{m \log (\delta^{-1})}{n\res}$ dominates the quantization error $m\delta$. Thus, we can bound the minimum over $\Rate \in \Lambda$ in \eqref{eq:pMLE_bound} from above by 
\begin{align*}
\min_{1 \le m \le \inDim} \left[ m^{-2\alpha} + \frac{m\log(\delta^{-1})}{n\res}\right] \approx \left(\frac{\log(\delta^{-1})}{n\res}\right)^{\frac{2\alpha}{2\alpha+1}}.
\end{align*}
Using $\tO(\cdot)$ notation to hide factors that are logarithmic in
$\inDim$ and $k$, we can particularize Theorem~\ref{thm:pMLE} to the
heavy-tail case:

\begin{theorem}\label{thm:compressible_pMLE}
\begin{align*}
  & \sup_{\Rate^\star \in \Sigma_{\alpha,L_0}} \Risk{\hRate^\pMLE_n}{\Rate^\star}  \nonumber\\
  & \quad = O(k^{-\alpha}) + \tO\left(\sqrt{\gamma k}\right) +
  \tO\left(\sqrt{\frac{k}{\gamma}}
    \left(\frac{\log(\delta^{-1})}{n\res}\right)^{\frac{\alpha}{2\alpha+1}}\right).
\end{align*}
\end{theorem}
Note that the risk bound here is worse than the benchmark bound of Theorem~\ref{thm:direct}. However, in order to compute the direct estimator one has to solve a linear program, whereas, as we show next, the pMLE can be approximated very efficiently with proper preprocessing of the observed counts $\Cts_n$ based on the structure of $\Graph_{k,\inDim}$.

\section{Efficient $\pMLE$ Approximation}
\label{sec:alg}
In this section we present an efficient algorithm for approximating the $\pMLE$ estimate. The algorithm consists of two phases: (1) first, we preprocess $\Cts_n$ to isolate a subset $\Vin_1$ of $\Vin = \{1,\ldots,\inDim\}$ which is sufficiently small and is guaranteed to contain the locations of the $k$ largest entries of $\Rate^\star$ (the whales); (2) then we construct a set $\Lambda$ of candidate estimators whose support sets lie in $\Vin_1$, together with an appropriate penalty, and perform pMLE over this reduced set.

The success of this approach hinges on the assumption that the magnitude of the smallest whale is much larger compared to the total contribution of the minnows. Specifically, we make the following assumption: Let $S \subset \Vin$ contain the locations of the $k$ largest coordinates of $\Rate^\star$. Then we require that
\begin{align}\label{eq:SNR_assumption}
\frac{\min_{i \in S} \rate^\star_i}{d} > \kerr{\Rate^\star}.
\end{align}
One way to think about \eqref{eq:SNR_assumption} is in terms of a signal-to-noise ratio, which must be strictly larger than the left degree $d$ of the underlying expander [recall that $d = O(\log(\inDim/k))$]. We also perturb our expander a bit as follows: choose an integer $k' > 0$ so that
\begin{align}\label{eq:expander_perturb}
\frac{15 k'd}{16} \ge kd + 1.
\end{align}
Then we replace our original $(k,1/16)$-expander with left-degree $d$ with a $(k',1/16)$-expander with the same left degree. The resulting procedure, displayed below as Algorithm~\ref{alg1}, has the following guarantees:


\begin{algorithm}[ht]
\caption{Efficient $\pMLE$ approximation algorithm}
\textbf{Input:}  Measurement vector $\Cts_\T$, and the sensing matrix $\AdjMat$.
\textbf{Output:}  An approximation $\has$
\begin{algorithmic}
\label{alg1}
\STATE Let $\Vout_1$ consist of the locations of the $kd$ largest elements of $\Cts_\T$ and let $\Vout_2 = \Vout \backslash \Vout_1$.
\STATE Let $\Vin_2$ contain the set of all variable nodes that have at least one neighbor in $\Vout_2$ and let $\Vin_1\nobreak= \Vin \backslash \Vin_2$. 
\STATE Construct a candidate set of estimators $\Lambda$ with support in $\Vin_1$ and a penalty $\pen(\cdot)$ over $\Lambda$.
\STATE  \label{finalstep}Output $ \arg\min_{\a\in  \Lambda}  \left[ - \log \Prob_{\T\tau \AdjMat \bd{\a}} (\Cts_\T) + 2 \pen(\bd{\a}) \right]$.
\end{algorithmic}
\end{algorithm}

\begin{theorem}
 \label{thm:sina}
The set $\Vin_1$ constructed by Algorithm~\ref{alg1} has the following properties: (1) $S \subset \Vin_1$; (2) $|\Vin_1| \le kd$; (3) $\Vin_1$ can be found in time $O(\inDim d) = O(\inDim \log(\inDim/k))$. 
 \end{theorem}
 \begin{IEEEproof}
(1) If we decompose $\Flow_n$ as $\Flow^S_n + \bd{e}$, then $\Cts_n = \AdjMat \Flow^S_n + \AdjMat \bd{e}$. Since each column of $\AdjMat$ is $d$-sparse and $\Flow^S_n$ is $k$-sparse, $\AdjMat \Flow^S_n$ is $kd$-sparse. On the other hand, $\Cts_n = \Cts^{\Vout_1}_n + \Cts^{\Vout_2}_n$, where, by construction, $\Cts^{\Vout_1}_n$ is the best $kd$-term approximation of $\Cts_n$. Hence,
\begin{align}\label{eq:eff_pMLE_1}
\| \Cts^{\Vout_2}_n \|_1 = \| \Cts_n - \Cts^{\Vout_1}_n \|_1 \le \| \AdjMat \bd{e} \|_1 \le d \| \bd{e} \|_1,
\end{align}
where the last inequality follows from the properties of $\AdjMat$. Now, since only the nodes in $\Vin_2$ have neighbors in $\Vout_2$,
\begin{align}
\| \Cts^{\Vout_2}_n \|_1 = \sum_{j \in \Vout_2} \sum_{i \in \Vin_2} A_{ji} \flow_{n,i} \ge \sum_{i \in \Vin_2} \flow_{n,i} = \| \Flow^{\Vin_2}_n \|_1.\label{eq:eff_pMLE_2}
\end{align}
Combining \eqref{eq:eff_pMLE_1} and \eqref{eq:eff_pMLE_2}, we get the bound $\| \Flow^{\Vin_2}_n \|_1 \le d \| \bd{e} \|_1$. Taking expectation of both sides, we obtain
\begin{align}\label{eq:eff_pMLE_3}
\Exp \| \Flow^{\Vin_2}_n \|_1 \le d \cdot \Exp \| \bd{e} \|_1 = d \cdot \Exp \kerr{\Flow_n} \le dn\res \kerr{\Rate^\star},
\end{align}
where the last step follows the same reasoning as in the proof of Theorem~\ref{thm:direct}. Now suppose that $S \cap \Vin_2 \neq \varnothing$. Then
\begin{align*}
\Exp \| \Flow^{\Vin_2}_n \|_1 \ge \Exp \| \Flow^{S \cap \Vin_2}_n \|_1 \ge n \res \min_{i \in S} \rate^\star_i > d n \res \kerr{\Rate^\star},
\end{align*}
where the last step follows from \eqref{eq:SNR_assumption}. Since \eqref{eq:eff_pMLE_3} must also hold, we arrive at a contradiction, and therefore $S \subset \Vin_1$.

(2) Suppose, to the contrary, that $|\Vin_1| > kd$. Let $\Vin'_1 \subseteq \Vin_1$ be any subset of size $kd+1$. Now, Lemma~3.6 in \cite{Khaj} states that, provided $\eps \le 1-1/d$, then every $(\ell,\eps)$-expander with left degree $d$ is also a $(\ell(1-\eps)d,1-1/d)$-expander with left degree $d$. We apply this result to our $(k',1/16)$-expander, where $k'$ satisfies \eqref{eq:expander_perturb}, to see that it is also a $(kd+1,1-1/d)$-expander. Therefore, for the set $\Vin'_1$ we must have $|\cN(\Vin'_1)| \ge  |\Vin'_1| = kd+1$. On the other hand, $\cN(\Vin'_1) \subset \Vout_1$, so $|\cN(\Vin'_1)| \le kd$. This is a contradiction, hence we must have $|\Vin_1| \le kd$.

(3) Finding the sets $\Vout_1$ and $\Vout_2$ can be done in $O(\outDim \log \outDim)$ time by sorting $\Cts_\T$. The set $\Vin_1$ can then can be found in time $O(\inDim d)$, by sequentially eliminating all nodes connected to each node in $\Vin_2$.
 \end{IEEEproof}

\begin{figure*}[ht]
  \centering \subfigure[Probability of successful support recovery as
  a function of number of whales
  $\k$.]{\includegraphics[width=0.30\textwidth]{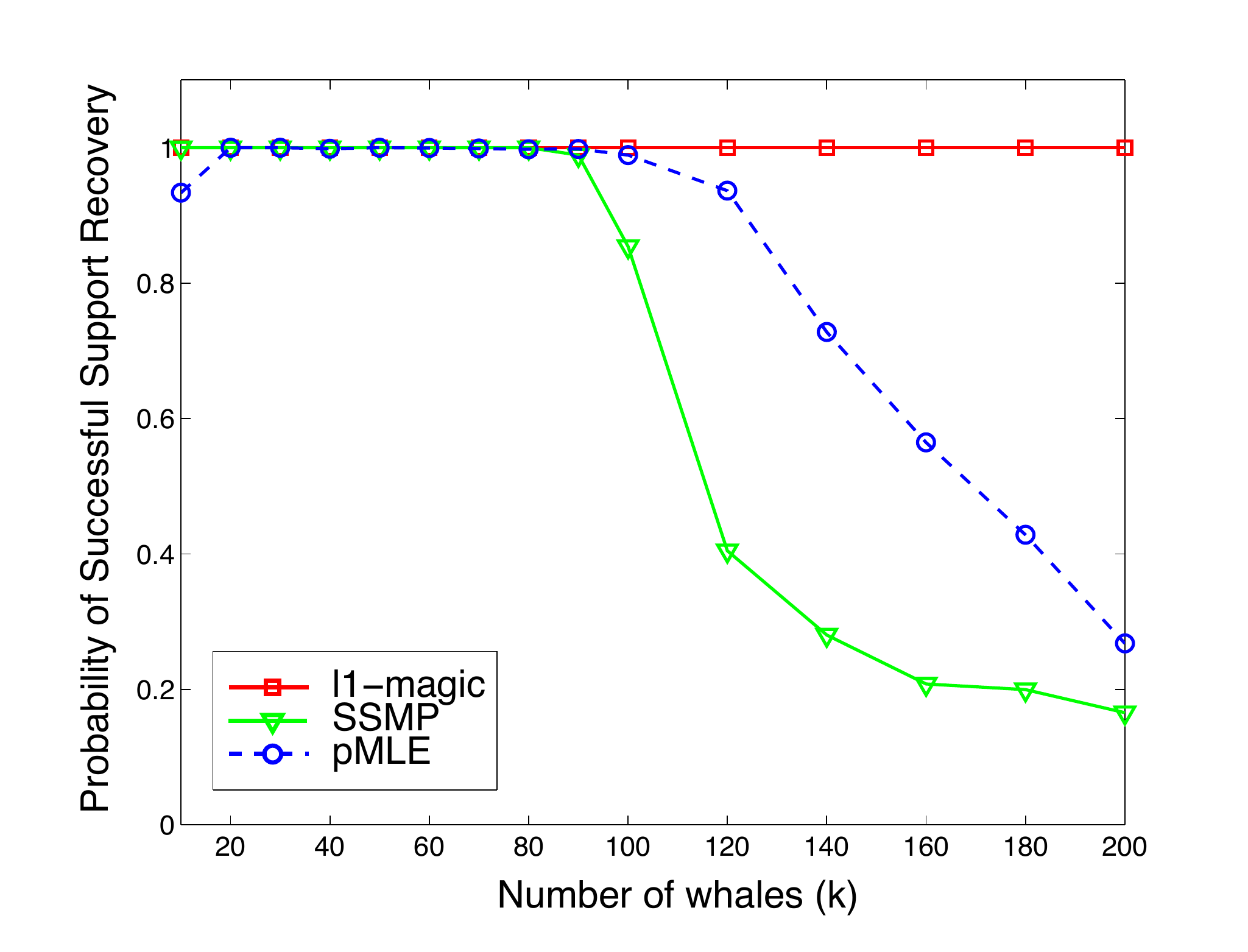}
    \label{fig1}} \subfigure[Relative $\ell_1$ error as a function of
number of whales $k$.]
{\includegraphics[width=0.30\textwidth]{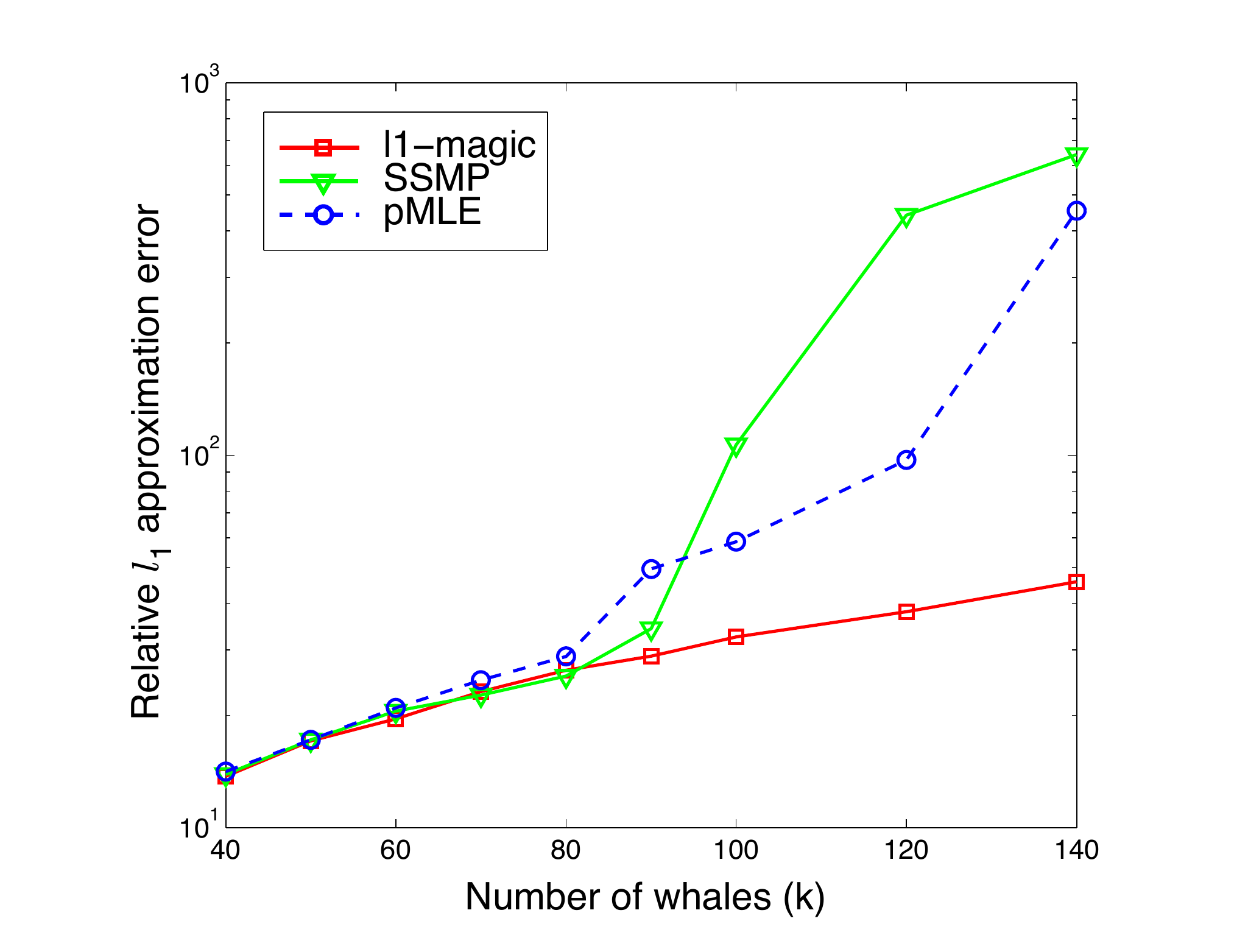}
\label{fig2}}
 \subfigure[Average recovery time as a function of  number of whales
$\k$.]{\includegraphics[width=0.30\textwidth]{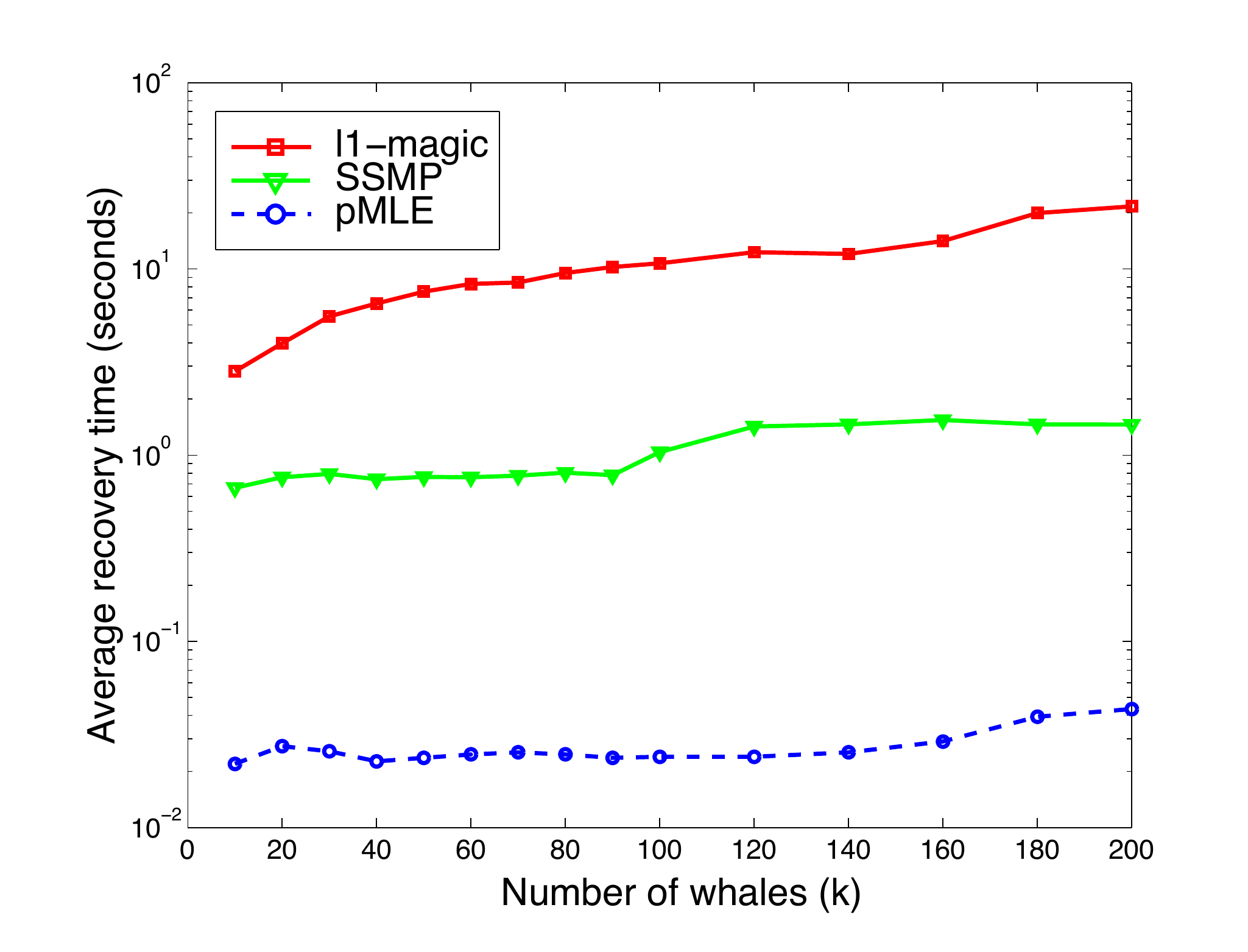}
\label{fig3}}
\caption{Performance - Complexity tradeoff for $\ell_1$-magic, SSMP
  and $\pMLE$. The number of flows $\n=5000$, the number of counters
  $\m=800$, and the number of updates $\T=40$. There are $k$ whales
  (peaks with magnitude 1), and the remaining entries are minnows with
  magnitudes determined by a ${\cal N}(0, 10^{-6})$ random variable.}
\label{fig1-3}
\end{figure*}
 
\begin{figure*}[ht]
  \centering \subfigure[Probability of successful support recovery as
  a function of number of whales
  $\k$.]{\includegraphics[width=0.30\textwidth]{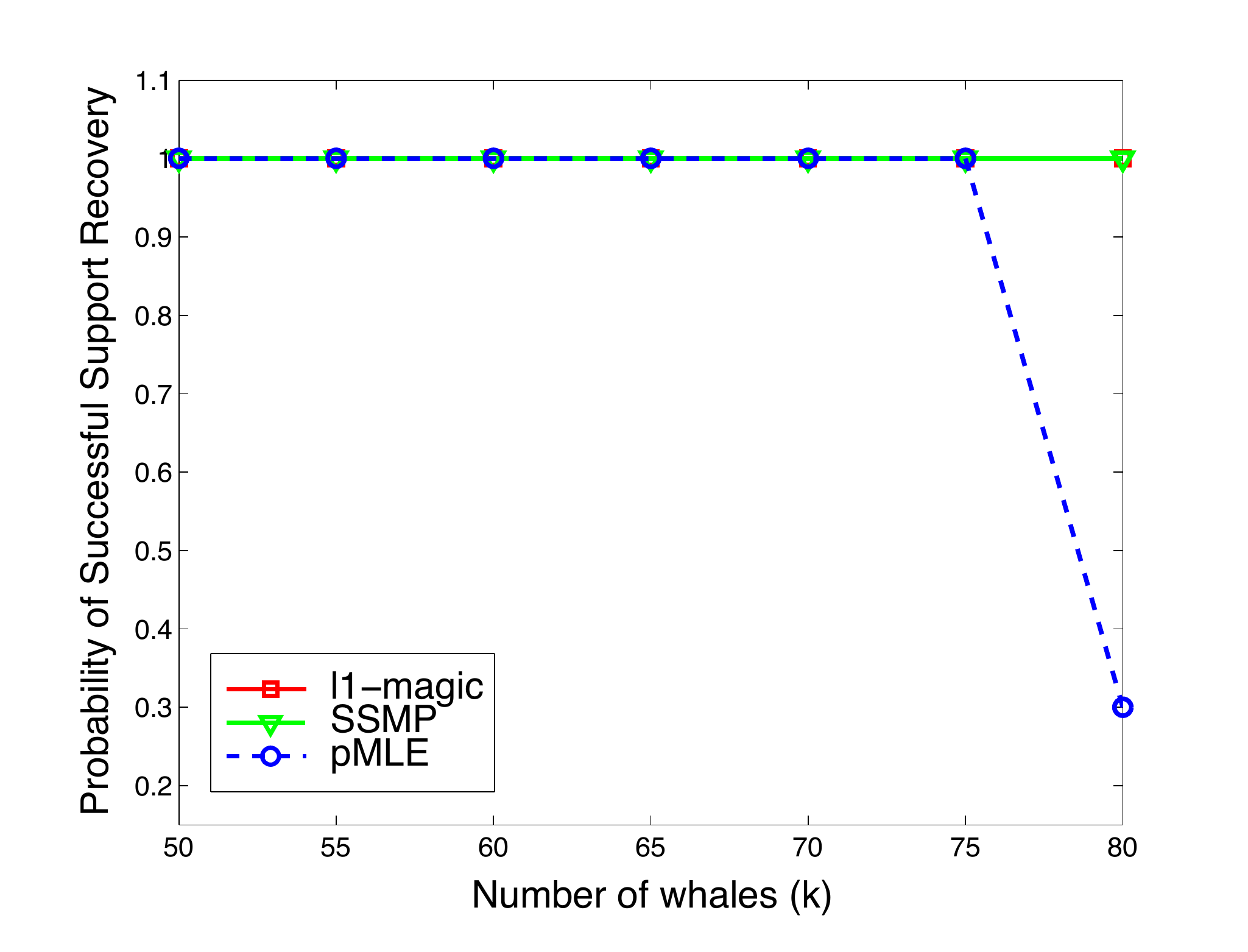}
\label{fig4}} 
\subfigure[Relative $\ell_1$ error as a function of
number of whales $k$.]
{\includegraphics[width=0.30\textwidth]{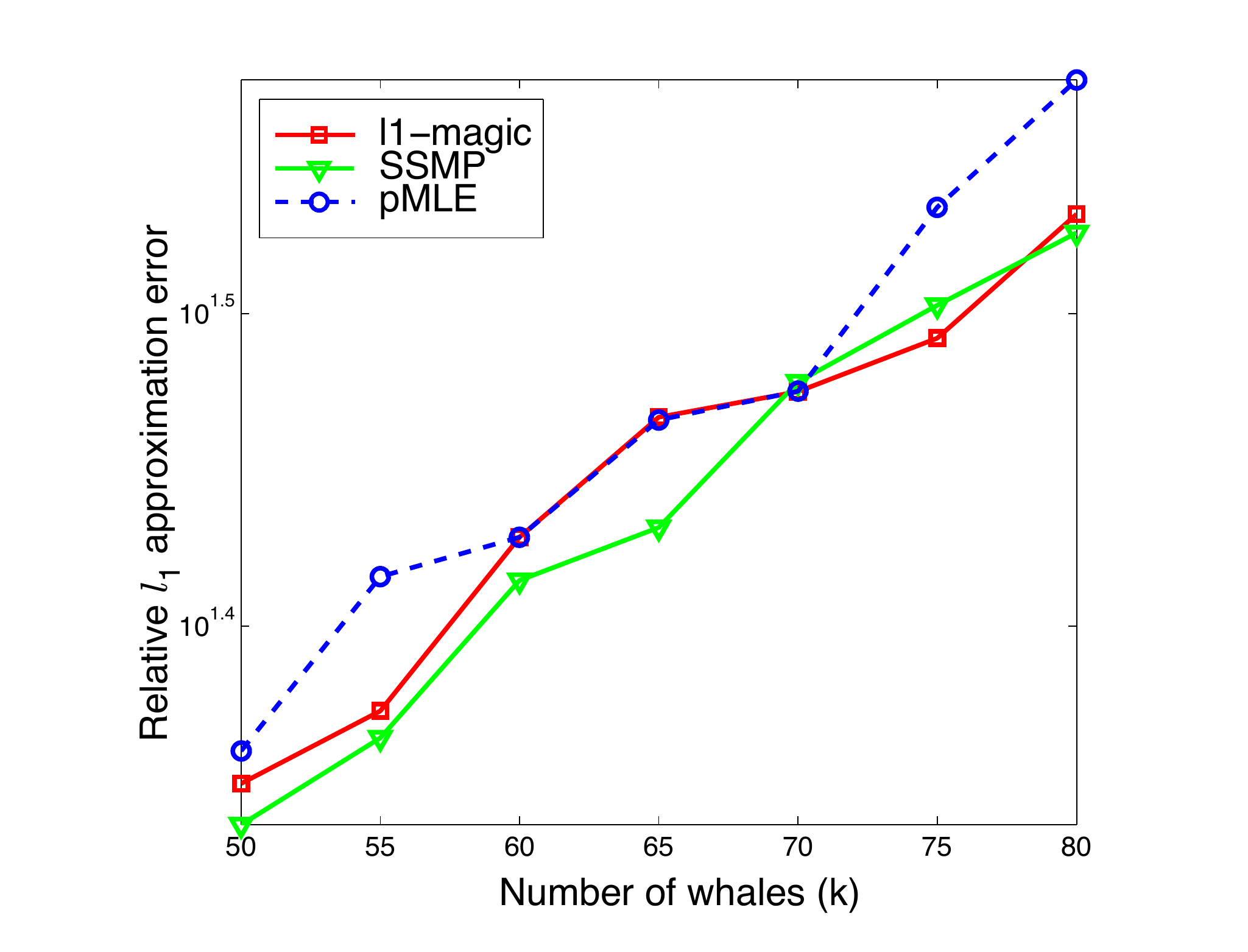}
\label{fig5}}
\subfigure[Average recovery time as a function of number of whales
$\k$.]{\includegraphics[width=0.30\textwidth]{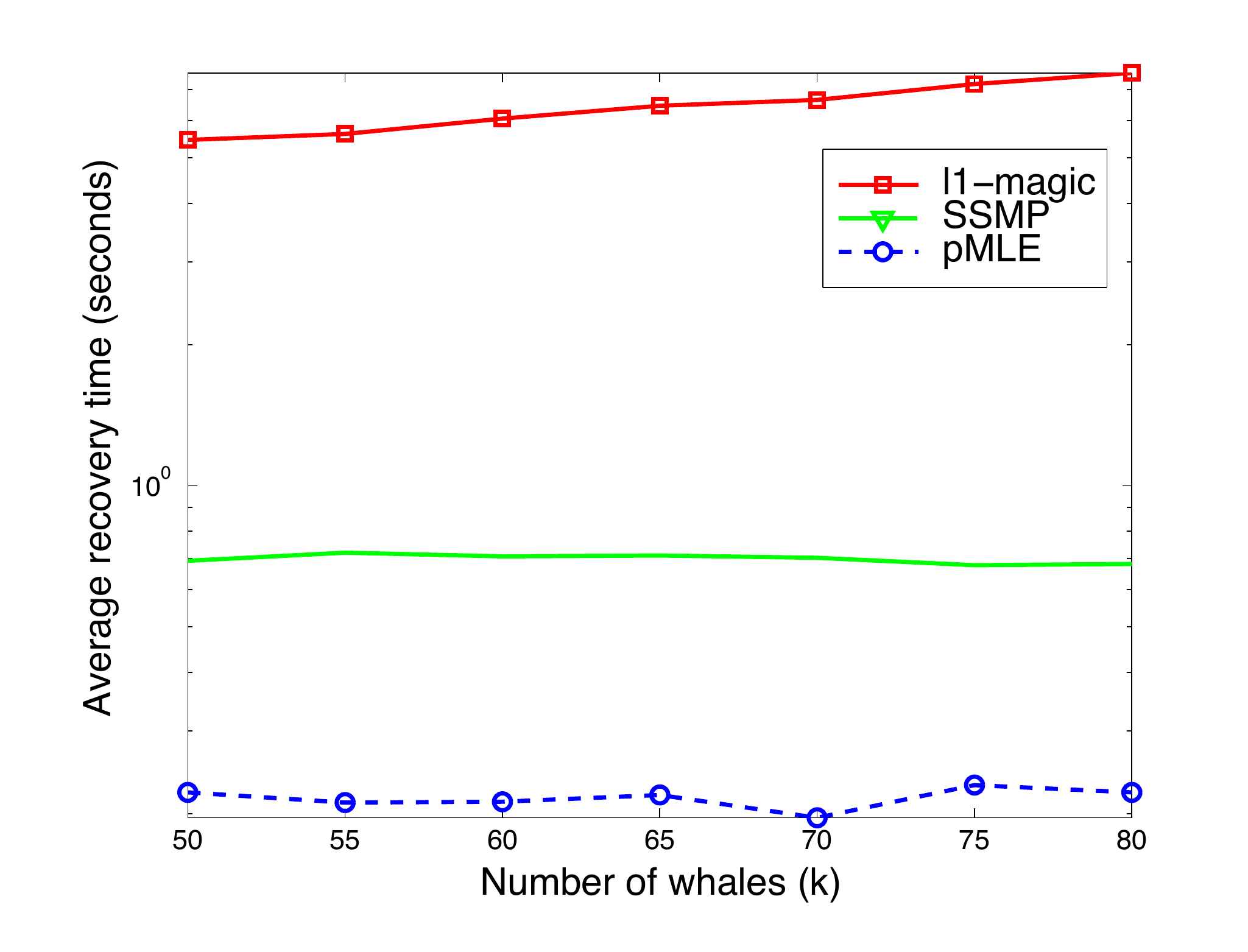}
\label{fig6}}
\caption{Performance - Complexity tradeoff for $\ell_1$-magic, SSMP
  and $\pMLE$. The number of flows $\n=5000$, the number of counters
  $\m=800$, and the number of updates $\T=40$. There are $k$ whales
  (peaks with magnitude determined by a ${\cal N}(0,1)$ random
  variable), and the remaining entries are minnows with magnitudes
  determined by a ${\cal N}(0, 10^{-6})$ random variable.}
\label{fig4-6}
\end{figure*}
  
 Having identified the set $\Vin_1$, we can reduce the pMLE optimization only to those candidates whose support sets lie in $\Vin_1$. More precisely, if we originally start with a sufficiently rich class of estimators $\tilde{\Lambda}$, then the new feasible set can be reduced to
 $$\Lambda \deq \left\{ \a\in\tilde{\Lambda} :  {\rm Supp}(\a)\subset \Vin_1  \right\}.$$ 
 Hence, by extracting the set $\Vin_1$, we can significantly reduce the complexity of finding the pMLE estimate. If $|\Lambda|$ is small, the optimization can be performed by brute-force search in $O(|\Lambda|)$ time. Otherwise, since $|\Vin_1| \le kd$, we can use the quantization technique from the preceding section with quantizer resolution $\sqrt{\delta}$ to construct a $\Lambda$ of size at most $(L_0/\sqrt{\delta})^{kd}$. In this case, we can even assign the uniform penalty
 $$
 \pen(\Rate) = \log |\Lambda|  = O\left( k \log(\inDim/k) \log(\delta^{-1})\right),
 $$
which amounts to a vanilla MLE over $\Lambda$.

\section{Experimental Results}
\label{sec:exp} 
Here we compare penalized MLE with $\ell_1$-magic \cite{l1}, a
universal $\ell_1$ minimization method, and with SSMP \cite{ssmp}, an
alternative method that employs combinatorial
optimization. $\ell_1$-magic and SSMP both compute the ``direct''
estimator by solving a convex program. The pMLE
estimate is computed using Algorithm~\ref{alg1} above and the Sparse
Poisson Intensity Reconstruction ALgorithm (SPIRAL) \cite{harmany:PCS}
for reconstruction of sparse signals from indirect Poisson
measurements.

Figures~\ref{fig1} through~\ref{fig6} report the result of numerical
experiments, where the goal is to identify the $k$ largest entries in
the rate vector from the measured data. The set of $k$ largest entries
(the whales) is chosen at random. Since a random graph is, with
overwhelming probability, an expander graph, each experiment was
repeated $30$ times \footnote{We observed similar results for
  experiments with larger number of trials.}.

Given a particular relative sizing of whales and minnows,
Figure~\ref{fig1} reports values of $k$ where recovery is possible
with generic $\ell_1$ algorithms ($\ell_1$-magic) but not with SSMP or
$\pMLE$. As $k$ increases the first algorithm to fail is SSMP, and the
probability of successful recovery falls more sharply than for
$\pMLE$.  We also report the relative $\ell_1$ error
($\|\Rate-\hRate_n\|_1/\|\Rate-\Rate^{(k)}\|_1$) as a function of $k$ in
Figure~\ref{fig2}.  However the complexity of $\ell_1$-magic is $2-3$
orders of magnitude greater than $\pMLE$, as shown in
Figure~\ref{fig3}.

The effect of increased variability in the size of whales is to reduce
the value of $\k$ at which $\pMLE$ fails. The size of minnows in
Figure~\ref{fig4-6} is the same as in Figure~\ref{fig1-3}, but the
variation in the size of whales is determined by an ${\cal N}(0,1)$
Gaussian random variable. Here we see that still Algorithm~\ref{alg1}
combined with SPIRAL is two order of magnitudes faster, but the
probability of success drops substantially for $k>80$.

\section{Conclusions}

The compressed sensing algorithms based on Poisson observations and
expander-graph sensing matrices provide a useful mechanism for
accurately and efficiently estimating a collection of flow rates with
relatively few counters. These techniques have the potential to
significantly reduce the cost of hardware required for flow rate
estimation. While previous approaches assumed packet counts matched
the flow rates exactly or that flow rates were i.i.d., the approach in
this paper accounts for the Poisson nature of packet counts with
relatively mild assumptions about the underlying flow rates
(i.e.,~that only a small fraction of them are large).

The ``direct'' estimation method (in which first the vector of flow
counts is estimated using a linear program, and then the underlying
flow rates are estimated using Poisson maximum likelihood) is
juxtaposed with an ``indirect'' method (in which the flow rates are
estimated in one pass from the compressive Poisson measurements using
penalized likelihood estimation). The direct method can yield smaller
error bounds, but this comes at a high computational cost relative to
the efficient algorithms associated with the indirect method. These
theoretical results are verified in our simulations.

The methods in this paper, along with related results in this area, are designed for settings in which the flow rates are sufficiently stationary, so that they can be accurately estimated in a fixed time window. Future directions include extending these approaches to a more realistic setting in which the flow rates evolve over time. In this case, the time window over which packets should be counted may be relatively short, but this can be mitigated by exploiting estimates of the flow rates in earlier time windows. Another direction for future research will be to tighten the bounds for the indirect method using oracle inequalities based on the Kullback--Leibler divergence.

\section*{Acknowledgment}

The work of M.~Raginsky and R.~Willett is supported by NSF CAREER Award No. CCF-06-43947, DARPA Grant No. HR0011-07-1-003, and NSF Grant DMS-08-11062. The work of R.~Calderbank and S.~Jafarpour is supported in part by NSF
under grant DMS 0701226, by ONR under grant N00173-06-1-G006, and by
AFOSR under grant FA9550-05-1-0443.

\bibliography{Poisson_flows.bbl}

\end{document}